\documentclass[aps,prl,showpacs,amsmath,amssymb,amsfonts,lengthcheck,twocolumn,longbibliography,superscriptaddress]{revtex4-2}

\usepackage{graphicx}
\usepackage{dsfont}
\usepackage{subfigure}
\usepackage{verbatim}
\usepackage{dcolumn}
\usepackage{bm}
\usepackage{epsf}
\usepackage{color}
\usepackage[toc]{appendix}
\usepackage{stmaryrd}
\usepackage{amsthm}
\usepackage{hhline}
\usepackage{amssymb}
\usepackage{mwe,tikz}\usepackage[percent]{overpic}
\usepackage{tikz}
\usetikzlibrary{arrows,shapes,calc,matrix}

\expandafter\let\csname equation*\endcsname\relax
\expandafter\let\csname endequation*\endcsname\relax
\usepackage{amsmath}
\usepackage{xstring}

\makeatletter
\newcommand\footnoteref[1]{\protected@xdef\@thefnmark{\ref{#1}}\@footnotemark}
\makeatother

\newcommand\myeqref[1]{
	Eq. (\textup{\ref{#1}})
}

\newcommand{\ptr}[2]{\mathrm{tr_{#1}}\left\{#2\right\}}

\newcommand{\bra}[1]    {\langle #1|}
\newcommand{\ket}[1]    {| #1 \rangle}
\newcommand{\bk}[2]     {\langle #1 | #2 \rangle}

\newcommand{\cS}        {{\mathcal S}}

\newcommand{\cE}        {{\mathcal E}}

\newcommand\cF{{\mathcal F}}
\newcommand\hocom[1]{}

\newcommand{\ba}{\begin{eqnarray}}
\newcommand{\ea}{\end{eqnarray}}
\newcommand{\bmath}{\begin{mathletters}}
\newcommand{\emath}{\end{mathletters}}
\newcommand{\ban}{\begin{eqnarray*}}
\newcommand{\ean}{\end{eqnarray*}}

\newcommand{\tr}[1]{\mathrm{tr}\left\{#1\right\}}

\newcommand{\bla}{bla\\bla\\bla\\bla\\bla}
\newcommand{\PRA}{Phys. Rev. A }

\usepackage{fmtcount}

\newcommand{\mc}[1]{\mathcal{#1}}

\newcommand{\draftmode}{1}    
\newcommand{\notetoself}[1]{\ifnum \draftmode=1 {\color[rgb]{0,0,0.8} [#1]} \fi}  
\newcommand{\cuttext}[1]{\ifnum \draftmode=1 {\color[rgb]{0,0.5,0} [#1]} \fi}  
\newcommand{\warntext}[1]{\ifnum \draftmode=1 {\color[rgb]{0.9,0.6,0} #1} \else {#1} \color{black} \fi}
\newcommand{\aref}[1]{{Appendix~\hyperref[#1]{A}}}
\newcommand{\bref}[1]{{Appendix~\hyperref[#1]{B}}}
\newcommand{\dref}[1]{{Appendix~\hyperref[#1]{C}}}
\usepackage[colorlinks=true,citecolor=blue,linkcolor=blue,urlcolor=blue]{hyperref}
\usepackage{cleveref}

\begin{document}

\title{Eavesdropping on the Decohering Environment:\\ Quantum Darwinism, Amplification, and the Origin of Objective Classical Reality}

\author{Akram Touil}
\email{akramt1@umbc.edu}
\affiliation{Department of Physics, University of Maryland, Baltimore County, Baltimore, MD 21250, USA}
\affiliation{Center for Nonlinear Studies, Los Alamos National Laboratory, Los Alamos, New Mexico 87545}

\author{Bin Yan}
\affiliation{Center for Nonlinear Studies, Los Alamos National Laboratory, Los Alamos, New Mexico 87545}
\affiliation{Theoretical Division, Los Alamos National Laboratory, Los Alamos, New Mexico 87545}

\author{Davide Girolami}
\affiliation{Politecnico di Torino, Corso Duca degli Abruzzi 24, Torino, 10129, Italy}

\author{Sebastian Deffner}
\affiliation{Department of Physics, University of Maryland, Baltimore County, Baltimore, MD 21250, USA}
\affiliation{Instituto de F\'isica `Gleb Wataghin', Universidade Estadual de Campinas, 13083-859, Campinas, S\~{a}o Paulo, Brazil}

\author{Wojciech Hubert Zurek}
\affiliation{Theoretical Division, Los Alamos National Laboratory, Los Alamos, New Mexico 87545}

\begin{abstract} 
``How much information about a system $\cS$ can one extract from a fragment $\cF$ of the environment $\cE$ that decohered it?'' is the central question of Quantum Darwinism. To date, most answers relied on the quantum mutual information of $\cS\cF$, or on the Holevo bound on the channel capacity of $\mathcal{F}$ to communicate the classical information encoded in $\mathcal{S}$. These are reasonable upper bounds on what is really needed but much harder to calculate -- the accessible information in the fragment $\cF$ about $\cS$. We consider a model based on imperfect {\tt c-not} gates where all the above can be computed, and discuss its implications for the emergence of objective classical reality. We find that all relevant quantities, such as the quantum mutual information as well as various bounds on the accessible information exhibit similar behavior. In the regime relevant for the emergence of objective classical reality this includes scaling independent of the quality 
of the imperfect {\tt c-not} gates or the size of $\cE$, and even nearly independent of the initial state of $\cS$. 
\end{abstract}

\maketitle

Quantum Darwinism \cite{Zurek2000AP,Zurek2003RMP,Ollivier2004PRL,Ollivier2005PRA,Zurek2009NP} explains the emergence of objective classical reality in our quantum Universe: The decohering environment $\cal E$ is a ``witness’’  who monitors and can reveal the state of the system $\cS$. Agents like us never measure systems of interest directly. Rather, we accesses fragments $\cal F$ of $\cal E$ that carry information about them. Since its inception \cite{Zurek2000AP}, Quantum Darwinism has advanced on both theory \cite{Giorgi2015PRA,Balaneskovic2015EPJD,Balaneskovic2016EPJD,Knott2018PRL,Milazzo2019PRA,Campbell2019PRA,Ryan2020,Garcia2020PRR,Lorenzo2020PRR,Qdc1,Qdc2,Qdc3,Qdc4,Qdc5,Qdc6,Qdc7,Qdc8,Qdc9,Qdc10,Qdc11} and experimental fronts \cite{Ciampini2018PRA,Chen2019SB,Unden2019PRL,Garcia2020NPJQI}.

Quantum mutual information $I({\cal S} : {\cal F})$ between an environment fragment and the system yields an upper bound on what $\cF$ can reveal about $\cS$. It has been used to estimate the capacity of the environment as a communication channel. We analyze a solvable model based on imperfect tunable {\tt c-not} (or {\tt c-maybe}) gates that couple $\cS$ to the subsystems of $\cE$. We compute the mutual information $I({\cal S}: {\cal E})$ as well as the Holevo $\chi({\cal S} : {\cal F})$  \cite{Holevo,nielsen2002quantum} -- that characterize the accessible information
in our {\tt c-maybe} - based model. We also compute the quantum discord \cite{Zurek2000AP,Ollivier2001PRL,Henderson2001JPA,Giorda2010PRL,Shi2011JPA,Zwolak2013SR,ZRZ2016SR} -- the difference of $I({\cal S} : {\cal F})$ and $\chi({\cal S} : {\cal F})$ that quantifies the genuinely quantum correlations between $\mc{S}$ and $\mc{F}$ \cite{Brodutch2011JPCS,Adesso2016,Bera2017RPP,discordc1}.



We find that $I({\cal S} : {\cal F})$ and $\chi({\cal S}, {\cal F})$ exhibit strikingly similar dependence on the size of $\cF$, with the initial steep rise followed by the classical plateau where, at the level set by the entropy $H_\cS$ of the system, the information $\cF$ has about $\cS$ saturates: Enlarging $\cF$ 
only confirms what is already known.  This behavior is universal and nearly independent of the initial state of $\mc{S}$ and the size of $\cE$.

\paragraph*{The model.}

The system $\mc{S}$ is a qubit coupled to $N$ independent non-interacting qubits of the environment  $\mc{E}$ via a {\tt c-maybe} gate,
\begin{equation}
U_{\oslash}=\begin{pmatrix}
1 & 0 & 0 & 0 \\
0 & 1 & 0 & 0 \\
0 & 0 & s & c \\
0 & 0 & c & -s
\end{pmatrix}.
\label{ope}
\end{equation}
The parameters $c=\cos(a)$ and $s=\sin(a)$ (where $a$ is the rotation angle of the target qubit) 
quantify the imperfection.

Our Quantum Universe $\cS\cE$ starts in a pure state:
\begin{equation}
\ket{\Psi_\mathcal{SE}^0} = \left(\sqrt{p}\, \ket { 0_\mathcal{S}} + \sqrt{q}\, \ket { 1_\mathcal{S}} \right) \bigotimes_{i=1}^{N} \ket {0^{i}} ,
\end{equation}
where $p+q=1$. The unitary $U_{\oslash}$ correlates each qubit in $\cE$ with $\mc{S}$,  and we obtain a branching state \cite{blume2005simple},
\begin{equation}
\ket{ \Psi^{\oslash}_\mathcal{SE}} = \sqrt{p}\,  \ket {0_\mathcal{S}} \bigotimes_{i=1}^{N} \ket { 0_{\cE_{i}}} + \sqrt{q}\,  \ket{1 _\mathcal{S}} \bigotimes_{i=1}^{N} \ket { 1_{\mathcal{E}_i}} \, .
\label{bstate}
\end{equation}
By construction $\ket{0_\mathcal{S}}$ and $\ket{1_\mathcal{S}}$ are the pointer states \cite{basis1,basis2}. They are orthogonal and immune to decoherence. The corresponding record states of $\mc{E}$ are
\begin{equation}
\ket { 0_{\cE_{i}}}  \equiv \ket{0^i}\quad\text{and}\quad \ket { 1_{\mathcal{E}_i}} \equiv s \ket{0^i}+ c \ket{1^i}\,,
\end{equation}
in terms of the orthogonal basis $\ket{0^i}$ and $\ket{1^i}$ of the $i$th qubit that defines $U_\oslash$, so that $\bk { 0_{\cE_{i}}} { 1_{\mathcal{E}_i}}= s$.

We will be interested in the correlations between the fragment $\cal F$ and $\cal S$. The marginal states of ${\cal S}$, an $m$-qubit fragment ${\cal F}_{m}$, and a bipartition ${\cal SF}_m$ are rank-two density matrices~\footnote{Deducing the reduced quantum states $\rho_\mathcal{S}$ and $\rho_{\mathcal{S}\mathcal{F}_m}$ in Eqs.~\eqref{eq:rho_S} and \eqref{eq:rho_SF} is straightforward, whereas $\rho_{\mathcal{F}_m}$ is slightly more involved, see e.g. \cite{ZQZ10}.}:
\begin{equation}
\label{eq:rho_S}
\rho_\mathcal{S}\equiv\ptr{\mc{E}}{\ket{ \Psi^{\oslash}_\mathcal{SE}}\bra{\Psi^{\oslash}_\mathcal{SE}} }=
\begin{pmatrix}
p & s^{N}\sqrt{pq}  \\
s^{N}\sqrt{pq}  & q 
\end{pmatrix} ,
\end{equation}

\begin{equation}
\label{eq:rho_F}
\rho_{\mathcal{F}_m}=
\begin{pmatrix}
p & s^{m}\sqrt{pq}  \\
s^{m}\sqrt{pq}  & q
\end{pmatrix},
\end{equation} 

\begin{equation}
\label{eq:rho_SF}
\rho_{\mathcal{S}\mathcal{F}_m}=
\begin{pmatrix}
p & s^{N-m}\sqrt{pq}  \\
s^{N-m}\sqrt{pq}  & q
\end{pmatrix}\,.
\end{equation}

{\it Symmetric quantum mutual information}
is often used to estimate the accessible information in $\cF$ in
 Quantum Darwinism \cite{Zurek2003RMP,Ollivier2004PRL,Ollivier2005PRA,blume2006quantum,Riedel2010PRL,Riedel2011NJP,Campbell2019PRA,Ryan2020,Garcia2020PRR}. It is defined using the von Neumann entropy, $H(\rho)=-\tr{\rho \log_{2}(\rho)}$ as;
\begin{equation}
{I}(\mc{S}:\mc{F}_m)=H_{\mc{S}}+H_{\mc{F}_m}-H_{\mc{S},\mc{F}_m}.
\label{MUTI}
\end{equation}
Joint entropy $H_{\mc{S},\mc{F}_m}$ quantifies the ignorance about the state of ${\mc{S}\mc{F}_m}$ in the tensor product of the Hilbert spaces of $\cS$ and $\cF$. 

In our model ${I}(\mc{S}:\mc{F}_m)$ can be computed exactly \cite{ZQZ10};
\begin{equation}
{I}(\mc{S}:\mc{F}_m)={h}(\lambda^{+}_{N,p})+{h}(\lambda^{+}_{m,p})-{h}(\lambda^{+}_{N-m,p}),
\label{mutinfo}
\end{equation}
where ${h}(x)=-x \log_2(x)-(1-x) \log_2(1-x)$ and $\lambda^{\pm}_{k,p}$ are the eigenvalues of the density matrices \eqref{eq:rho_S} -- \eqref{eq:rho_SF},
\begin{equation}
\lambda^{\pm}_{k,p}=\frac{1}{2}\left(1 \pm \sqrt{\left(q-p\right)^{2}+4s^{2k}pq}\right)\,.
\label{root}
\end{equation}
We thus have a closed expression for the mutual information ${I}(\mc{S}:\mc{F}_m)$.

\begin{figure}
	\includegraphics[width=0.482\textwidth]{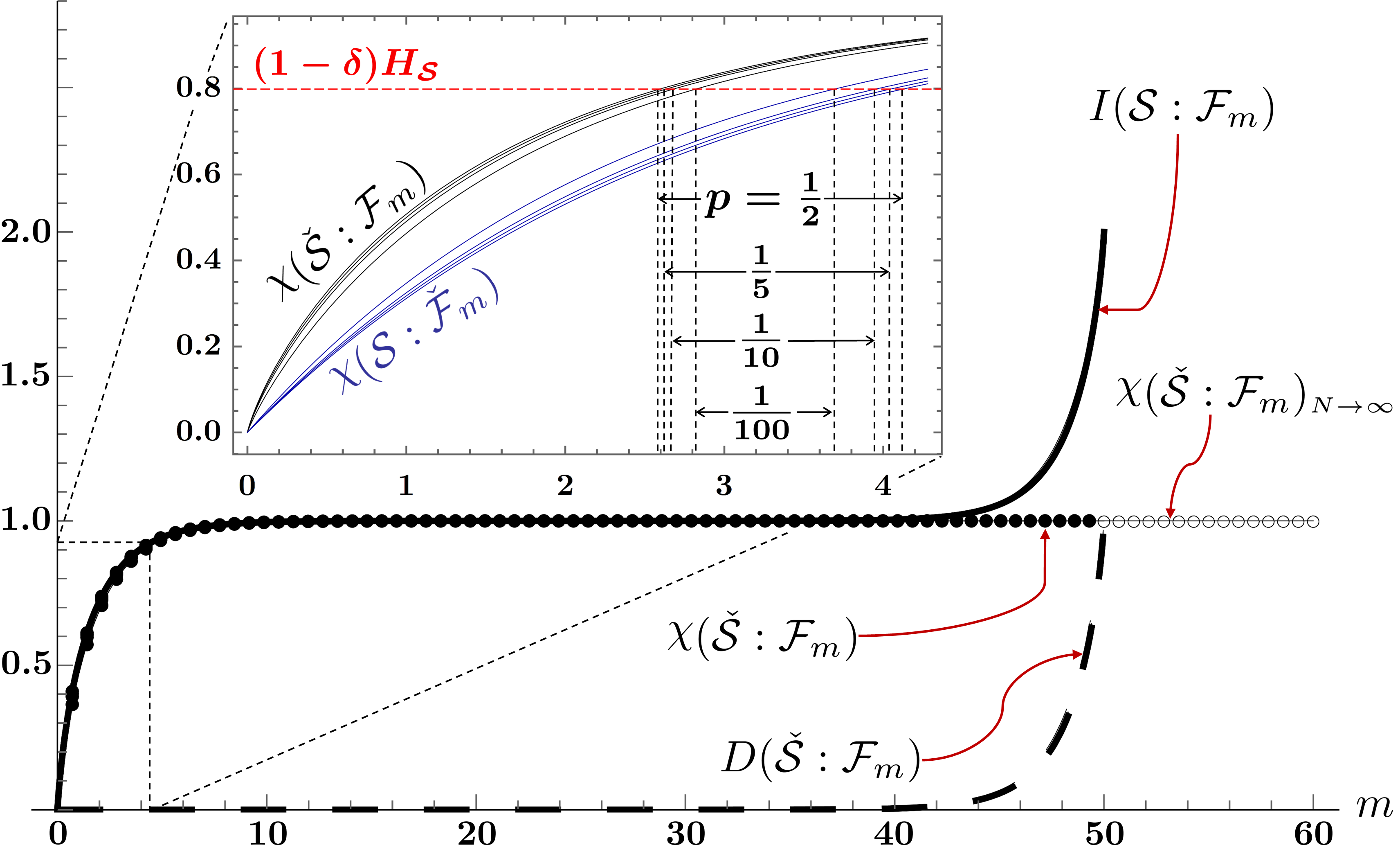}
	\caption{\label{fig:mutI}{\it Approximate universality of mutual information:} Symmetric $I(\cS : \cF_m)$ and Holevo bound $\chi(\check \cS : \cF_m)$ coincide until the fragment $\cF_m$ becomes almost as large as $\cE$. Renormalized $I(\cS : \cF)/H_\cS$ and $\chi(\check \cS : \cF)/H_\cS$ depend only weakly on the probabilities of the outcomes (see inset). Their difference -- quantum discord $D(\check \cS : \cF)$ -- vanishes until $\cF_m$ begins to encompass almost all of $\cE$, $m \sim N/R_\delta$. The inset also compares the ratios $\chi(\check \cS : \cF)/H_\cS$ and $\chi( \cS : \check \cF)/H_\cS$ computed for several probabilities $p$ of the pointer state $\ket {0_\cS}$ in Eq.~\eqref{bstate}. Note that the fragment sizes $m_\delta$ that supply $\sim$80\% of information about $\cS$ are only modestly affected by $p$ and quite similar for these two different information measures.} 
\end{figure}

As seen in Fig.~\ref{fig:mutI}, symmetric mutual information ${I}(\mc{S}:\mc{F}_m)$ exhibits a steep initial rise with increasing fragment size $m$, as a larger $\cF_m$ provides more data about $\cS$. This initial rise is followed by a long \emph{classical plateau}, where the additional information imprinted on the environment is redundant.  

Note that, when $\cS \cE$ is in a pure state,  the entropy of a fragment $\cF$ is equal to $H_{\cS d\cF}$, that is the entropy $\cS$ would have if it was decohered only by the fragment $\cF$.  When we further assume {\it good decoherence} \cite{blume2005simple,ZQZ10} -- i.e., that the off-diagonal terms of $\rho_\cS$ and $\rho_{\cS \cF_m} $ are negligible (which in our model corresponds to $s^{N-m} \ll  s^m$) -- we obtain an approximate equality; 
\begin{equation}
I(\cS : \cF_m) = H_{\cF_m} = H_{\cS d \cF_m}, 
\label{gooddeco}
\end{equation}
since $H_\cS=H_{\cS \cF_m}$ cancel one another in Eq. ~\eqref{MUTI}.  Furthermore,  when the environment fragments are typical~\cite{CoverThomas} (and in our model all fragments of the same size are identical -- hence, each is typical) the plot of ${I}(\mc{S}:\mc{F}_m)$ is antisymmetric around ${I}(\mc{S}:\mc{F}_m) = H_{\cal S}$ and $m=N/2$ \cite{blume2005simple}.  

We will see that the behavior of  ${I}(\mc{S}:\mc{F}_m)$ is approximately universal. This means that after suitable re-scaling its functional form is nearly independent of the size of the environment $N$, of the quality of the {\tt c-maybe} gate $U_\oslash$, and almost independent of the initial state of $\cS$.

Agents generally do not insist on knowing the state of $\cS$ completely, but tolerate a finite {\it information deficit} $\delta$. When $I(\cS : \cF_{m_\delta}) \geq (1-\delta) H_\cS$ is attained already for a fragment with $m_\delta \ll N$ subsystems, a fraction $f_\delta=m_\delta /N$ of the environment, then there are many ($1/f_\delta$) such fragments. We define redundancy of the information about $\cS$ in $\cE$ via:
\begin{equation}
\label{eq:redun}
{\cal R}_{\delta} \equiv N/m_\delta \quad \text{with}\quad I(\cS : \cF_{m_\delta}) = (1-\delta) H_\cS\,.
\end{equation}
Redundancy ${\cal R}_{\delta}$ is the length of the classical plateau in the units set by $m_\delta$, see Fig.~\eqref{fig:mutI}. The beginning of the plateau is determined by the smallest $m_\delta$ such that $I(\cS : \cF_{m_\delta}) \geq (1-\delta) H_\cS$. 

In realistic models $I(\cS : \cF)=H_\cS$ only when $ f = 1/2$ (see \cite{blume2005simple}). Thus, significant redundancy appears only when the requirement of completeness of the information about $\cS$ that can be extracted from $\cF$ is relaxed. Moreover,  Eq.~\eqref{eq:redun} is an overestimate since $I(\cS : \cF_{m_\delta}) $ is only an upper bound of what can be found out about $\cS$ from $\cF$ \footnote{Note that $H_{\cS}$ -- in the circumstances of interest to us -- is not the thermodynamic entropy of $\cS$. Rather, it is the missing information about the few {\it relevant} degrees of freedom of $\cS$. The thermodynamic entropy of a cat, for instance, will vastly exceed the information an observer is most interested in -- e.g., the one crucial bit in the `diabolical contraption' envisaged by Schr{\"o}dinger.}.

We will now consider better estimates: Inset in Fig.~\ref{fig:mutI} compares $I(\cS : \cF_{m_\delta}) $ with the two Holevo - like $\chi$'s we are about to discuss and illustrates resulting fragment sizes (hence, redundancies) they imply.



\paragraph*{Asymmetric mutual information} 
is defined using conditional entropy. We mark the system whose states are used for such conditioning by an inverted ``hat'', so when it is $\check{\mc{S}}$ we consider the asymmetric mutual information:
\begin{equation}
J(\check \cS : \mc{F}_m)_{\{ \ket {s_k} \}}  = H_{\cF} - H_{{\cF}|\check \cS_{\{ \ket {s_k} \}}} \ .
\label{Asmutinfo}
\end{equation}
Above, $H_{{\cF_m} | {\check \cS_{\{ \ket {s_k} \}}}}$  is the conditional entropy \cite{nielsen2002quantum} that quantifies the missing information about $\cF$ remaining after the observable with the eigenstates $\{ \ket {s_k} \}$ was measured. Accordingly, the joint entropy in Eq.~\eqref{MUTI} is replaced by;
\begin{equation}
H_{{\cF_m}, \check \cS_{\{ \ket {s_k}\}}}= H_{{\cF_m} | {\check \cS_{\{ \ket {s_k} \}}}} + H_{\check \cS_{\{ \ket {s_k}\}}}\,.
\label{CondS}
\end{equation}
The asymmetric joint entropy  depends on whether $\cS$ or $\cF$ are measured and on the measurements that are used. The entropy increase associated with the wavepacket reduction means that the asymmetric entropy \eqref{CondS} is typically larger than the symmetric version $H_{\cS, {\cF_m}} $ in Eq.~\eqref{MUTI}: Local measurements cannot extract information encoded in the quantum correlations between $\mc{S}$ and $\mc{F}_m$, which is why the asymmetric $J(\check \cS : \mc{F}_m)$ is needed, \cite{nielsen2002quantum}; see also \cite{touil2021quantum}. 

For  optimal  measurements the asymmetric $J(\check \cS : \mc{F}_m)$ defines the Holevo bound \cite{Holevo},
\begin{equation}
\label{eq:holevo}
J(\check \cS:\mc{F}_m)=\max_{\{ \ket {s_k} \}} J(\check \cS:\mc{F}_m)_{\{ \ket {s_k} \}} \equiv \chi (\check \cS:\mc{F}_m) .
\end{equation}
In our model,
measurement of the pointer observable of $\cS$ is optimal~\cite{ZQZ10}. Indeed,~\myeqref{bstate} shows that in the pointer basis $\{\ket {0_\cS}, \ket {1_\cS}\}$ the conditional entropy disappears, $H_{{\cF_m} | {\check \cS}} = 0$, as states of $\cF_m$ correlated with pointer states of $\cS$ are pure.

The limit of large $\cE$ ($N \ge N-m \gg m$) reflects typical situation of agents (who do not even know the size of $\cE$, and only access ``their ${\cal F}_m$'', with $m \ll N$). This is  good decoherence, $s^N \le s^{N-m} \ll s^m$, and equations simplify:  
\hocom{As already noted, $H_{\cS,\cF_m} = H_{\cS}$.
Therefore;
\ba
I(\mc{S}:\mc{F}_m) \approx H_{\mc{F}_m}={h}(\lambda^{+}_{m,p}) = 
J(\check \cS : \mc{F}_m)=
\chi(\check \cS : \mc{F}_m) . \ \ \ \ \ \ 
\label{Mutinfo}
\ea
The equality $J(\check{\mc{S}}:\mc{F}_m)\approx I(\mc{S}:\mc{F}_m)$ assumes optimal measurements on $\check \cS$. 
In case of good decoherence
measurements of the pointer states are optimal \cite{ZQZ10}.

Indeed, for model and for good decoherence \cite{blume2005simple,ZQZ10} measuring the pointer observable of $\cS$ is actually optimal \cite{ZQZ10}. Observe in Eq.~\eqref{bstate} that in the pointer basis $\{\ket {0_\cS}, \ket {1_\cS}\}$ the conditional entropy disappears, $H_{{\cF_m} | {\check \cS}} = 0$, since all states of $\cF_m$ correlated with the pointer states of $\cS$ are pure. Moreover, good decoherence, $s^N \le s^{N-m} \ll s^m$, is equivalent to large environments, $N \ge N-m \gg m$. 
}
Using $H_{\cS,\cF_m} = H_{\cS}$ and Eq.~\eqref{gooddeco} we can thus write,
\begin{equation}
I(\mc{S}:\mc{F}_m) \approx H_{\mc{F}_m}={h}(\lambda^{+}_{m,p}) =\chi(\check \cS : \mc{F}_m). 
\label{Mutinfo}
\end{equation}
An immediate important consequence is that  $H_{\cF_m}$ determines both the symmetric $I(\mc{S}:\mc{F}_m)$ (except for the final rise) as well as the asymmetric (optimal) $J(\check \cS : \mc{F}_m)=\chi(\check \cS : \mc{F}_m)$. We have:
\begin{widetext}
\begin{equation}
\chi(\check \cS : \mc{F}_m)=-\frac{1}{2}\,\log _{2}\left( p q\left(1-s^{2 m}\right) \right)-\sqrt{1-4p q\left(1-s^{2 m}\right)}\, \operatorname{Arctanh}_{2}\left(\sqrt{1-4p q\left(1-s^{2 m}\right)}\right)\,,
\label{ChiS}
\end{equation}
\end{widetext}
where ``$\operatorname{Arctanh}_2$'' denotes $\operatorname{Arctanh}/\ln(2)$.

Fig.~\ref{fig:mutI} compares  $\chi(\check{\mc{S}}:\mc{F}_m)$ with ${I}(\mc{S}:\mc{F}_m)$ for finite  and infinite $N$ and for different values of $s$ and $p$. 
As it shows, Eq. (\ref{ChiS}) matches $I(\mc{S}:\mc{F}_m)$ until the far end ($N-m \ll m$) of the classical plateau. This is a consequence of two scalings: (i) ``vertically'', the plateau appears at $H_\cS = -p\log_2( p) - q \log_2 (q)$, and it is easy to see that for  $s^N \ll s^m \ll 1$ we have $\chi(\check{\mc{S}}:\mc{F}_m)=H_\cS$ in Eq.~\eqref{ChiS}; (ii) ``horizontally",  $H_{\mc{F}_m}$ depends on $s^m$, so weakly entangling gates can be compensated by using more of them -- larger $m$. What is surprising is how insensitive are these plots to $p$, the probability of the outcome.

This remarkably universal behavior is a consequence of \emph{good decoherence} \cite{ZQZ10}.  Both,  $\rho_{\mathcal{S}}$ and $\rho_{\mathcal{S}\mathcal{F}_m}$, Eqs.~\eqref{eq:rho_S} and \eqref{eq:rho_SF}, become diagonal in the pointer basis. Moreover, the quality of $U_{\oslash}$ (set by $c$ and $s$) determines the `` information flow rate'' from $\cal S$ to $\cal F$. Thus, when (at a fixed $p$) one demands the same $H_{\mc{F}_m}$, this translates into identical $\rho_{\cF_m}$  when $s_1^{m_1} \simeq s_2^{m_2}$. Therefore, less efficiently entangling gates can be compensated by relying on more of them -- on a suitably enlarged $\cF$, with $m_2 = m_1 \,\log(s_1)/\log(s_2)$. 

\paragraph{Environment as a communication channel.}
While the mutual information $I({\mc{S}}:\mc{F}_m)$ is easier to compute and a safe upper bound on the accessible information in $\cF_m$, it is important to verify it is also a reasonable estimate of that accessible information (as generally assumed in much of the Quantum Darwinism literature). 
The asymmetric mutual information extracted by optimal measurements on the environment fragment $\cF_m$ is:
\begin{equation}
 J(\cS : \check \cF_m) =  H_\cS - H_{\cS|\check{\cF}_m} = \chi(\cS : \check \cF_m)\,.
 \label{JF}
\end{equation}
The joint entropy given in terms of the conditional entropy $H_{\cS | {\check \cF}_m}$ now becomes,
\begin{equation}
H_{\cS, \check \cF_m} = H_{\cS | {\check \cF}_m} + H_{{\check \cF}_m}\, .
\label{CondF}
\end{equation}
As in Eq.~\eqref{CondS} above, all terms in  Eq.~\eqref{CondF} depend on how $\cF$ is measured. However, while measuring $\mc{S}$ in the pointer basis simplified the analysis (since e.g. $H_{{\cF}|{\check \cS}_{\{ \ket {s_k}\}}}=0$) this is no longer the case when $\cF_m$ is measured. 

To compute $\chi(\cS : \check \cF_m)$ we rely on the Koashi-Winter monogamy relation~\cite{KW}. Details of that calculation are relegated to the supplementary material~\cite{SM}.

\begin{figure}
	\includegraphics[width=0.482\textwidth]{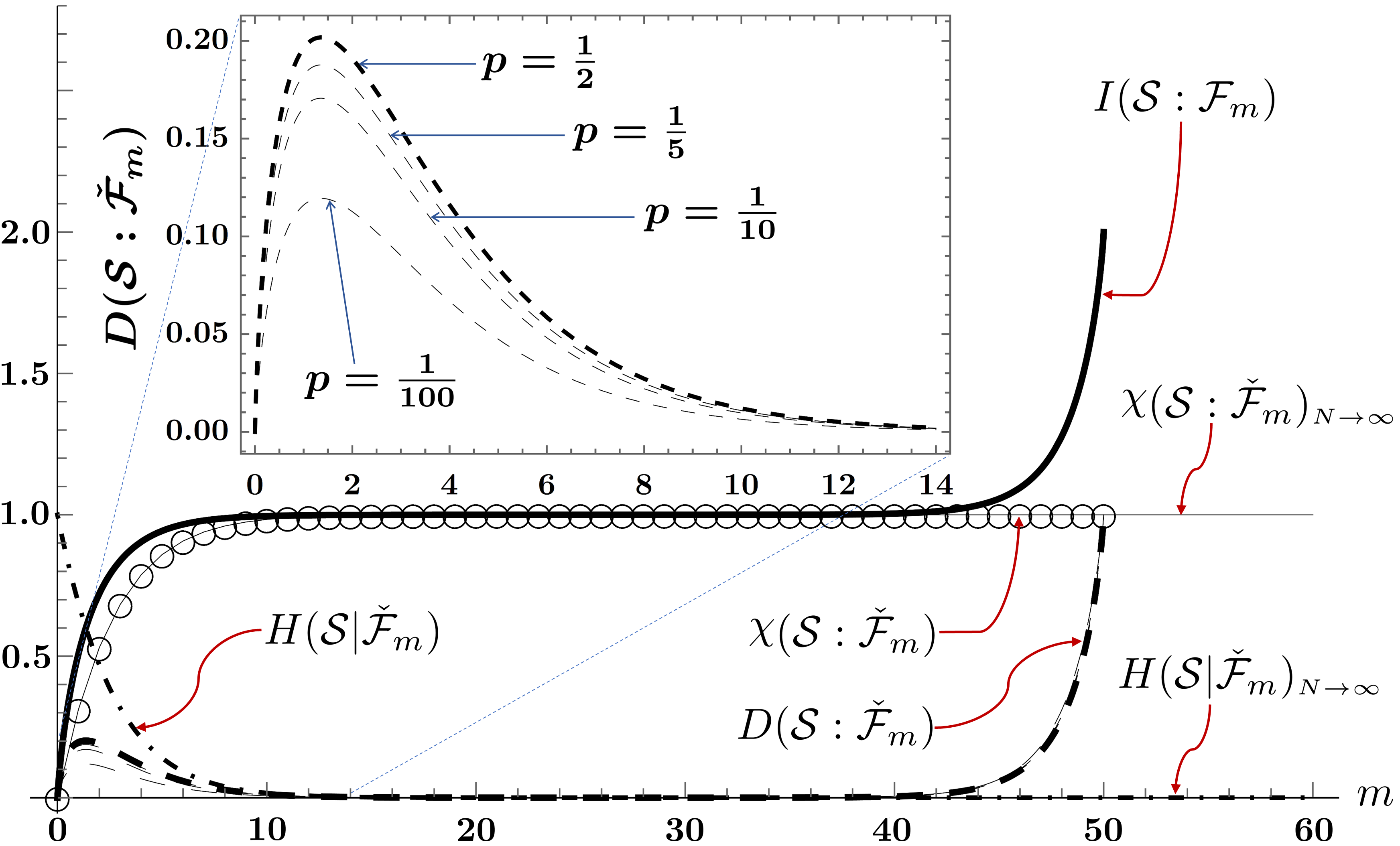}
	\caption{\label{fig:D}{\it Accessible information in the environment fragment.} Fragment $\cF_m$ carries at most $\chi(\cS : \check \cF)$ of classical information about the system it helped decohere. As seen above, this Holevo - like quantity is less than the symmetric mutual information $I(\cS : \cF)$ or the Holevo bound $\chi(\check{\mc{S}}:\mc{F}_m)$. Their difference (quantum discord $D(\cS : \check \cF)$) is significant already early on (in contrast to $D(\check \cS :  \cF)$), but disappears as the plateau is reached. It reappears again (as did $D(\check \cS :  \cF)$) when $\cF_m$ begins to encompass almost all of $\cE$. }
\end{figure}

We focus again on the limit of large $\cE$ ($N \ge N-m \gg m$): Agents only access ``their ${\cal F}_m$'', a small fraction of $\cE$ with $m \ll N$.  Assuming good decoherence we obtain
\begin{widetext}
\begin{equation}
\label{eq:holevo_F}
\chi(\mc{S}:\check \cF_m)= H_\cS +\frac{1}{2}\, \log_{2}\left(pqs^{2m}\right) +\sqrt{1-4pqs^{2m}}\,\operatorname{Arctanh}_2\left(\sqrt{1-4pqs^{2m}}\right)\,.
\end{equation}
\end{widetext}
Equation~\eqref{eq:holevo_F} constitutes our main result. We have decomposed the Holevo-like quantity $\chi(\mc{S}:\check \cF_m)$ into the plateau entropy $H_\cS$ and $H_{\cS|{\check \cF_m}}$ -- the ignorance about $\cS$ remaining in spite of the optimal measurement on $\cF_m$~\cite{elaborate}. Rather remarkably,  $H_{\cS|{\check \cF_m}}= H_\cS - \chi(\mc{S}: \check {\mc{F}}_m) $ -- the conditional entropy -- scales {\it exactly} with $pqs^{2m}$. What remains to do is to quantify the differences of $I({\mc{S}}:\mc{F}_m)$ and  $\chi (\check \cS:\mc{F}_m)$ with  $\chi(\mc{S}:\check \cF_m)$. In Fig.~\ref{fig:D} we compare it with these other, easier to compute, quantities. 

{\it Redundancy} of the information about $\cS$ in the channel $\cF_m$ can be now computed using $\chi(\mc{S}:\check \cF_m)$, Eq. \eqref{eq:holevo_F}, and compared with the estimates based on $I(\cS:\cF_m)$. The fragment $\cF_{m_\delta}$ can carry all but the deficit $\delta$ of the classical information about the pointer state of $\cS$ when $\chi(\mc{S}:\check \cF_{m_\delta})\ge(1-\delta) H_\cS$. This leads to a transcendental equation for $m_\delta$ that we solve numerically: $R_\delta = N / m_{\delta}$, where $N$ is the number of subsystems in $\cE$.

The inset in Fig.~\ref{fig:mutI} shows that -- while $m_\delta$ deduced using $I(\cS:\cF_m)\approx \chi (\check \cS:\mc{F}_m)$ do not coincide with those obtained using $\chi(\mc{S}:\check \cF_m)$ -- the difference is modest, unlikely to materially affect conclusions about the emergence of objective classical reality. Indeed, in the supplementary materials we estimate that the redundancy estimates based on $I(\cS:\cF_m)$ and $\chi(\mc{S}:\check \cF_m)$ differ at most by $\sim 37\%$ for $\delta \leq 0.2$,
and by much less in the regime where $\delta \rightarrow 0$.

In situations relevant for observers who rely on photons, $R_{\delta=0.1} \simeq 10^8$ is amassed when sunlight illuminates a $1\mu m$ dust grain in a superposition with a $1\mu m$ spatial separation for $1 \mu s$ ~\cite{Riedel2010PRL,Riedel2011NJP}. It may seem like we are stretching the applicability of our {\tt c-maybe} model too far, but the equations for $I({\mc{S}}:\mc{F}_m)$ and $\chi (\check \cS:\mc{F}_m)$ derived for photon scattering {\it coincide } with our Eq. \eqref{ChiS}, see supplement~\cite{SM}. Thus, it appears that the information transfer from $\cS$ to $\cE$ leading to the buildup of redundancy has universal features captured by our model.

{\it Quantum discord} is the difference between symmetric \eqref{mutinfo} and asymmetric quantum mutual information \cite{Ollivier2001PRL,Henderson2001JPA,Giorda2010PRL,Shi2011JPA,Zwolak2013SR,ZRZ2016SR,Brodutch2011JPCS,Adesso2016,Bera2017RPP,discordc1}. The \emph{systemic discord} is defined as;
\begin{equation}
D(\check{\mc{S}}:\mc{F}_m) = I({\mc{S}}:\mc{F}_m)  - \chi (\check \cS:\mc{F}_m)  \,.
\label{Discord}
\end{equation}
The measurements on pointer observables of $\mc{S}$ are optimal. 

Mutual information for pure decoherence induced by non-interacting subsystems of $\cE$ can be written as \cite{ZQZ10,Z07b}:
\begin{equation}
I(\cS:\cF)=\stackrel {local/classical} {\bigl(H_\cF-H_\cF(0)\bigr)} + \stackrel {global/quantum} {\bigl(H_{\cS d \cE} - H_{\cS d \cE_{\backslash\cF}}\bigr)}\,.
\label{4.20}
\end{equation}
As $\cS\cE$ is a pure product state, the initial entropy of $\mc{F}$ is zero, $H_\cF(0)=0$. Assuming good decoherence and conditioning on the pointer basis (hence, 
$\chi(\check \cS :\mc{F}_m)=H_{\mc{F}_m}$), Eq. (\ref{Mutinfo}) we have
\begin{equation}
I({\mc{S}}:{\mc{F}}_m) - J(\check \cS : {\mc{F}}_m) = H_{\cS d \cE} - H_{\cS d \cE_{\backslash{\cF}_m}},
\label{discord}
\end{equation}
where $H_{\cS d \cE}$ ($H_{\cS d \cE_{\backslash\cF}}$) is the entropy of the system \textit{decohered} by $\cE$ (or just by $\cE_{\backslash\cF}$ -- i.e., $\cE$ less the fragment $\cF$).   

The global/quantum term represents quantum discord  in the pointer basis of $\cS$ \cite{ZQZ10}. Good decoherence implies $H_{\cS d \cE} \approx H_{\cS d \cE_{\backslash\cF}}$, so $D(\check{\mc{S}}:\mc{F}_m) \approx 0$. As long as $\cE_{\backslash\cF}$ is large enough to induce good decoherence, Eq. (\ref{Mutinfo}) holds, and, hence,  the systemic discord \eqref{Discord} vanishes~\footnote{Some have advocated ``strong Quantum Darwinism'' \cite{LeCastro} where the Holevo information  $\chi (\check \cS:\mc{F}_m)$ based on the measurement of $\cS$ (rather than $I({\mc{S}}:\mc{F}_m)$) would play a key role.  At least in case of good decoherence, and in view of Eq. ~\eqref{Mutinfo} which implies $\chi (\check \cS: \mc{F}_m) \approx I({\mc{S}}:\mc{F}_m)$ when $m \ll N - m<N$, such distinctions appear unnecessary.}.

Systemic quantum discord can become large again when $\cF_m$ encompasses almost all $\cE$, as in this case $H_{\cS \cF_m}$ approaches $H_{\cS \cE}=0$ (given our assumption of a pure $\cS \cE$). In this (unphysical) limit $I(\cS : \cF_m)$ climbs to $H_{\cF_m} + H_\cS = 2 H_{\cS}$, while $\chi (\check \cS:\mc{F}_m) \leq H_{\cF_m}$. As good decoherence implies $\chi (\check \cS:{\mc{F}}_m) \approx H_{\cF_m}$, $D(\check {\mc{S}}:\mc{F}_m)$ can reach $H_{\cF_m}$. Indeed, when $\cS\cE$ is pure, $\chi (\check \cS: \mc{F}_m)$ and $D(\check{\mc{S}}:\mc{F}_m)$ -- classical and quantum content of the correlation -- are complementary~\cite{Zwolak2013SR},  see Fig.~\ref{fig:mutI}.

The \emph{fragmentary discord} is the difference between the mutual information $I({\mc{S}}:\mc{F})$ and what can be extracted from $\mc{SF}$ by measuring only the fragment $\cF$:
\ba
I({\mc{S}}:\mc{F})  - \chi (\cS: \check \cF) \approx \chi (\check \cS:\mc{F}) - \chi(\cS : \check \cF).
\ea  
It can be evaluated:
\begin{equation}
\begin{split}
D(\cS : \check \cF_m) &\approx H_{\check \cF_m} - \left( H_\cS - H_{\cS|{\check \cF_m}}\right),\\ 
& = \left(H_{\cS|{\check \cF_m}}+H_{\check \cF_m}\right)- H_\cS\, .
\end{split}
\end{equation}
The bracketed terms in the last two expressions represent different quantities. The difference between the symmetric and asymmetric mutual information  $H_{\check \cF_m} - \left( H_\cS - H_{\cS|{\check \cF_m}}\right)$ is the original definition of discord.

Note that initially decoherence does not suppress fragmentary discord $D(\cS : \check \cF_m)$. This is because the states of $\cF_m$ that are correlated with the pointer states of $\cS$ are not orthogonal: The scalar product of the branch fragments $\cF_m$ corresponding to $\ket {0_\cS}$ and $\ket {1_\cS}$ is $s^m$. Thus, while the symmetric mutual information increases with $m$, orthogonality is approached gradually, also as $m$ increases.  Perfect distinguishability, i.e., orthogonality of record states of $\cF$ is needed to pass on all the information about $\cS$ \cite{Z07a, Z13,Gardas2016PRA}. See again Fig.~\ref{fig:D} for an illustration of these findings.

\paragraph*{Concluding remarks.}
We found that in the pre-plateau regime relevant for emergence of objective reality (where $I({\cal S} : {\cal F})$ increases with the size of $\cF$) the mutual information as well as the Holevo bound $\chi(\check {\cal S} : {\cal F})$ coincide and exhibit universal scaling behaviors independent of the size of $\cE$, of how imperfect are the {\tt c-maybe}'s, and only weakly dependent on the probabilities of pointer states. The corresponding Holevo $\chi(\check {\cal S} : {\cal F})$ and $I({\cal S} : {\cal F})$ coincide until $\cF$ encompasses almost all of $\cE$. 

However, the accessible information $\chi({\cal S} : \check {\cal F})$ in the environment fragments $\cF$ differs somewhat from $I({\cal S} : {\cal F})$ in the pre-plateau region. This difference tends to be small compared to, e.g., the level of the plateau, and disappears as the plateau is reached for larger fragments. This behavior -- generic when many copies of the information about $\cS$ are deposited in the environment -- facilitates estimates of the redundancy of the information about the system in the environment, as the differences between $I({\cal S} : {\cal F}) \approx \chi(\check {\cal S} : {\cal F})$ or $\chi({\cal S}: \check {\cal F})$ are noticeable but inconsequential.

To sum up, sensible measures of information flow lead to compatible conclusions about $R_\delta$. The differences in the estimates of redundancy based on these quantities are insignificant for the emergence of objective classical reality -- the overarching goal of Quantum Darwinism.  The functional dependence of the symmetric mutual information in the photon scattering model~\cite{Riedel2010PRL,Riedel2011NJP} is the same as in our model. Thus, the universality we noted in scaling with $s$ and $p$ (approximate for $I({\cal S} : {\cal F})=\chi (\check \cS: \cF_m) $, exact for $\chi (\cS: \check \cF_m) $) may be a common attribute of the information that reaches us, human observers.

\acknowledgements{We acknowledge several discussions with Michael Zwolak, who has provided us with extensive and perceptive comments that have greatly improved presentation of our results. This research was supported by grants FQXiRFP-1808 and FQXiRFP-2020-224322 from the Foundational Questions Institute and Fetzer Franklin Fund, a donor advised fund of Silicon Valley Community Foundation (SD and WHZ, respectively), as well as by the Department of Energy under the LDRD program in Los Alamos. A.T., B.Y. and W.H.Z. also acknowledge support from U.S. Department of Energy, Office of Science, Basic Energy Sciences, Materials Sciences and Engineering Division, Condensed Matter Theory Program, and the Center for Nonlinear Studies. D. G. acknowledges financial support from the Italian Ministry of Research and Education (MIUR), grant number 54$\_$AI20GD01.}

\bibliography{opm}

\begin{thebibliography}{65}
\bibitem [{ela()}]{elaborate}%
  \BibitemOpen
  \href@noop {} {\bibinfo  {journal} {Strictly speaking, Holevo bound $\chi$ on
  the channel capacity presumes conditioning on a predetermined classical basis
  (such as the pointer basis of $\cS$ in $\chi(\check S : \cF_m)$). In general,
  $\cF_m$ in $\chi(\mc{S}:\check \cF_m)$ does not possess such a predetermined
  classical basis, hence the ``Holevo-like'' designation.}\ }\BibitemShut
  {NoStop}%
\end{thebibliography}

\end{document}


\title{Supplemental Material\\
	 Eavesdropping on the Decohering Environment:\\ Quantum Darwinism, Amplification, and the Origin of Objective Classical Reality}

\author{Akram Touil}
\email{akramt1@umbc.edu}
\affiliation{Department of Physics, University of Maryland, Baltimore County, Baltimore, MD 21250, USA}
\affiliation{Center for Nonlinear Studies, Los Alamos National Laboratory, Los Alamos, New Mexico 87545}

\author{Bin Yan}
\affiliation{Center for Nonlinear Studies, Los Alamos National Laboratory, Los Alamos, New Mexico 87545}
\affiliation{Theoretical Division, Los Alamos National Laboratory, Los Alamos, New Mexico 87545}

\author{Davide Girolami}
\affiliation{Politecnico di Torino, Corso Duca degli Abruzzi 24, Torino, 10129, Italy}

\author{Sebastian Deffner}
\affiliation{Department of Physics, University of Maryland, Baltimore County, Baltimore, MD 21250, USA}
\affiliation{Instituto de F\'isica `Gleb Wataghin', Universidade Estadual de Campinas, 13083-859, Campinas, S\~{a}o Paulo, Brazil}

\author{Wojciech Hubert Zurek}
\affiliation{Theoretical Division, Los Alamos National Laboratory, Los Alamos, New Mexico 87545}
	

\maketitle


		
    
\setcounter{equation}{0}
\setcounter{figure}{0}
\setcounter{table}{0}
\setcounter{page}{1}
\makeatletter
\renewcommand{\theequation}{S\arabic{equation}}
\renewcommand{\thefigure}{S\arabic{figure}}
\renewcommand{\bibnumfmt}[1]{[R#1]}
\renewcommand{\citenumfont}[1]{R#1}
Within the framework of Quantum Darwinism, the emergence of classicality from the quantum substrate is a direct consequence of the redundant imprinting of classical information, about the system of interest $\mc{S}$, in different fragments $\mc{F}$ of the environment. This classical information, obtained by eavesdropping on environmental fragments, is upper bounded by the Holevo quantity $\rchi(\mc{S}:\check{\mc{F}})$ (aka the Holevo bound), which can be expressed as
\begin{equation}
\rchi(\cS : \check \cF)= H_\cS - H_{\cS|\check\cF},
\end{equation}
such that $H_i=-\trace{\rho_i\log_2(\rho_i)}$ stands for the von Neumann entropy, and $H_{\cS|\check\cF}$ is the conditional von Neumann entropy~\cite{nielsen2002quantum} that accounts for the missing information about the system $\cS$ after optimal measurements are applied on the fragment $\cF$.

In the present supplemental material, we provide the technical details leading to analytic expressions of $\rchi(\mc{S}:\check{\mc{F}})$, in the physical model we described in the manuscript~\cite{darwint1}. We separate the supplemental material into five sections. First, we present a general overview of the Koashi-Winter relation as the main tool of our derivation. We then show how it applies to our specific physical model of a central qubit undergoing decoherence in a many-qubit environment. The second part of the supplemental material details the necessary steps leading to analytic expressions of the Holevo bound (in the general case and in the limit of good decoherence). Using this result, we illustrate the statements made in the manuscript regarding the redundancy of information and the breaking of the universal behavior of both the mutual information and the Holevo bound. Finally, in the final section we explicitly explore the connection, through the quantum mutual information, between our many-qubit model and the photon scattering model. The latter connection shows that our analysis captures universal attributes in cases when decoherence is caused by environments composed of noninteracting subsystems such as photons.

\section{The Koashi-Winter relation}
For any tripartite system $ABC$, such that $\rho_{ABC}$ is pure, the Koashi-Winter relation can be written as~\cite{KW}
\begin{equation}
	\rchi(A:\check{B})+E(A:C)=H_A,
	\label{kw}
\end{equation}
where ``$E(A:C)$'' refers to the entanglement of formation~\cite{concurrence1,concurrence2,wootters2001entanglement} of $\rho_{AC}$, which is an entanglement measure for bipartite mixed quantum states (defined as an extension of the entanglement entropy to mixed states). To elaborate, the entanglement of formation ``$E(A:C)$'' quantifies the amount of entanglement in the mixed state $\rho_{AC}$, since for such cases (i.e. for mixed states) the entanglement entropy is no longer a viable measure of entanglement. The latter is due to the fact that the von Neumann entropy is not symmetric for general mixed states~\cite{nielsen2002quantum}.

The Koashi-Winter relation~(\ref{kw}) represents a monogamy relation between the entanglement of formation and the classical correlations in an extended Hilbert space. In other words, this relation captures the trade-off between entanglement and classical correlations. For a given value of the von Neumann entropy $H_{A}$, the higher the entanglement subsystem ``$A$'' shares with ``$C$'', the less classical information about ``$A$'' is accessible through optimal measurements on ``$B$''.

For two-qubit states $\rho_{AC}$, the entanglement of formation can be expressed as a function of the concurrence~\cite{wootters2001entanglement}
\begin{equation}
	E\left(A:C\right)=h\left(\frac{1+\sqrt{1-\left(\mathrm{Con}\left(A:C\right)\right)^{2}}}{2}\right),
\end{equation}
where ``$\mathrm{Con}\left(A:C\right)$'' stands for the concurrence of the state $\rho_{A C}$, and ${h}(x)=-x \log_2(x)-(1-x) \log_2(1-x)$. Furthermore, the entanglement measure~\cite{emeasure,wootters2001entanglement} ``$\mathrm{Con}\left(A:C\right)$'' can be analytically determined by computing the eigenvalues $\mu_i$ of the 4x4 density matrix $\rho_{AC}\tilde{\rho}_{AC}$, where $\tilde{\rho}_{AC}=\left(\sigma_y \otimes \sigma_y \right)  \rho^{*}_{AC} \left(\sigma_y \otimes \sigma_y \right)$. For two-qubit states where $\mu_i \geq \mu_{i+1}$, the analytic formula reads~\cite{concurrence2}
\begin{equation}
	\mathrm{Con}(A:C)=\max \left\{0, \sqrt{\mu_{1}}-\sqrt{\mu_{2}}-\sqrt{\mu_{3}}-\sqrt{\mu_{4}}\right\}.
	\label{con1}
\end{equation}

In the physical model of a central qubit $\cS$ undergoing decoherence in a many-qubit environment $\cE$, assuming that the state of the quantum universe $\mc{S}\mc{E}$ is pure implies that, by decomposing the environment into typical fragments, the state $\rho_{\mc{S}\mc{F}_m\mc{F}_{N-m}}$ is pure. Therefore, in this model, the Koashi-Winter relation is written as
\begin{equation}
	\rchi(\mc{S}:\check{\mc{F}}_m)+E(\mc{S}:\mc{F}_{N-m})=H_{\mc{S}}.
	\label{kwm}
\end{equation}
Based on the above expression, in order to determine the Holevo bound we need to evaluate the entanglement of formation. In fact, given the structure of the branching state of the dynamics~\cite{darwint1}
\begin{equation}
	| \Psi_\mathcal{SE} \rangle= \sqrt{p}  \ket { 0_\mathcal{S}} \bigotimes_{i=1}^{N} \ket { 0_{\cE_{i}}} + \sqrt{q} | 1 _\mathcal{S}\rangle \bigotimes_{i=1}^{N} \ket { 1_{\mathcal{E}_i}} \ ,
\end{equation}
tracing out the degrees of freedom of a given partition of the environment, or the system of interest, results in a density matrix of rank-two at most, i.e. the resulting density matrices are regarded as virtual qubits~\footnote{Here, the notion of virtual qubit is a mathematical construct where the corresponding states can be mapped to a Hilbert space of dimension equal to two.}. This implies that the bipartite system ``$\mc{S}\mc{F}_{N-m}$'' is a qubit-virtual qubit pair. Therefore, we have
\begin{equation}
	E\left(\mc{S}:\mc{F}_{N-m}\right)=h\left(\frac{1+\sqrt{1-\left(\mathrm{Con}\left(\mc{S}:\mc{F}_{N-m}\right)\right)^{2}}}{2}\right).
	\label{enof}
\end{equation}
Additionally, since we are dealing with rank-two (at most) density matrices, when computing the concurrence there are only two nonzero eigenvalues ($\mu_1$ and $\mu_2$), which implies that~\myeqref{con1} simplifies to
\begin{equation}
	\mathrm{Con}\left(\mc{S}:\mc{F}_{N-m}\right)=|\sqrt{\mu_1}-\sqrt{\mu_2}|.
	\label{con2}
\end{equation}

In summary, to determine the Holevo bound $\rchi(\mc{S}:\check{\mc{F}}_m)$ we first need to evaluate $\mu_1$ and $\mu_2$ from the expression of the 4x4 density matrix $\rho_{\mc{S}\mc{F}_{N-m}}$. From these eigenvalues we can determine the concurrence (cf.~\myeqref{con2}), which directly results in an analytic expression for the entanglement of formation (cf.~\myeqref{enof}). Thus, we can infer the general form of the Holevo bound $\rchi(\mc{S}:\check{\mc{F}}_m)$, and the corresponding expression in the good decoherence limit (presented in Equation (20) of the manuscript~\cite{darwint1}).

In what follows, we present the technical details that led to the expressions of $\mu_1$ and $\mu_2$. We start with instructive case of $m=1$, and then generalize to arbitrary values of $m$ such that $1\leq m < N$.

\section{The Holevo bound}
\paragraph{Single qubit of the environment $\mc{F}_1$:}
We split the branching state $| \Psi_\mathcal{SE} \rangle$ into two partitions. The first is composed of the central qubit $\mc{S}$ and a single qubit of $\mc{E}$, while the second partition is the rest of the environment. We adopt the following notations,
\begin{equation}
	\begin{split}
		| \boldsymbol{0} \rangle &\equiv \bigotimes_{i=1}^{N-1} | 0_{\cE_{i}} \rangle,\ \ | \boldsymbol{1} \rangle \equiv \frac{1}{\mc{N}}\left(\bigotimes_{i=1}^{N-1} |  1_{\cE_{i}} \rangle-s^{N-1}\bigotimes_{i=1}^{N-1} |  0_{\cE_{i}} \rangle\right),\\
		| \thickbar{0} \rangle &\equiv|00\rangle ,\ \ | \thickbar{1} \rangle \equiv s|10\rangle+c|11\rangle,
		\label{basis1}
	\end{split}
\end{equation}
with $\mc{N}=\sqrt{1-s^{2(N-1)}}$, and
\begin{equation}
(\forall i \in \llbracket 1, N\rrbracket); 
\ \	|  0_{\cE_{i}} \rangle \equiv | 0 \rangle ,\ | 1_{\cE_{i}} \rangle \equiv s | 0 \rangle + c | 1 \rangle.
\end{equation}
The real parameters $c$ and $s$, quantify the degree with which the environmental qubits monitor the central qubit $\cS$. From the above definitions, we have $\langle \boldsymbol{i}|\boldsymbol{j} \rangle=\delta_{i,j}$ and $\langle \thickbar{i}|\thickbar{j} \rangle=\delta_{i,j}$ for $(i,j) \in \{0,1\}^2$, and the state of the universe $\mc{S}\mc{E}$ can be written as
\begin{equation}
		| \Psi_\mathcal{SE} \rangle = \sqrt{p} |\thickbar{0}\boldsymbol{0}\rangle+\sqrt{q}s^{N-1}|\thickbar{1}\boldsymbol{0}\rangle+\sqrt{q}\sqrt{1-s^{2(N-1)}}|\thickbar{1}\boldsymbol{1}\rangle.
\end{equation}
Therefore, the rank-two density matrix representing the composite state of the central qubit $\mc{S}$ and a single qubit of the environment $\mc{F}_1$ is
\begin{equation}
	\rho_{\mc{S}\mc{F}_1}=\begin{pmatrix}
		p && 0 && s^{N}\sqrt{pq} && s^{N-1}c\sqrt{pq} \\
		0 && 0 && 0 && 0\\
		s^{N}\sqrt{pq}&& 0 && s^2 q && s c q\\
		s^{N-1}c\sqrt{pq}&& 0 && s c q && c^2 q
	\end{pmatrix}.
\end{equation}
Now we generalize to the case of an arbitrary environmental fragment $\cF_m$ (with $m$ qubits) in order to determine the density matrix $\rho_{\mc{S}\mc{F}_m}$, for any $1\leq m < N$, which directly implies the expression of $\rho_{\mc{S}\mc{F}_{N-m}}$ and the corresponding eigenvalues $\mu_1$ and $\mu_2$.

\paragraph{Environmental fragment $\mc{F}_m$:}
Considering a partition of the environment (with $m$ qubits), and similar to the previous single qubit analysis (cf.~\myeqref{basis1}), we group the central qubit with ``$m$'' qubits from the environment to get
\begin{equation}
	| \thickbar{0} \rangle \equiv|0\rangle \bigotimes_{i=1}^{m} | 0_{\cE_{i}} \rangle,\ | \thickbar{1} \rangle \equiv |1\rangle \bigotimes_{i=1}^{m} |  1_{\cE_{i}} \rangle,
\end{equation}
which in turn can be written as a basis for ``the qubit-virtual qubit'' pair
\begin{equation}
	| \thickbar{0} \rangle \equiv|0\mathfrak{0}\rangle ,\ | \thickbar{1} \rangle \equiv |1\rangle\left( s^{m}|\mathfrak{0}\rangle+\sqrt{1-s^{2m}}|\mathfrak{1}\rangle\right),
	\label{basism}
\end{equation}
with
\begin{equation}
	|\mathfrak{0} \rangle \equiv \bigotimes_{i=1}^{m} | 0_{\cE_{i}} \rangle,\  | \mathfrak{1} \rangle \equiv \frac{1}{\mc{M}}\left(\bigotimes_{i=1}^{m} |  1_{\cE_{i}} \rangle-s^{m}\bigotimes_{i=1}^{m} |  0_{\cE_{i}} \rangle\right),
\end{equation}
such that $\mc{M}=\sqrt{1-s^{2m}}$ and $\langle \mathfrak{i}|\mathfrak{j} \rangle=\delta_{\mathfrak{i},\mathfrak{j}}$. Following this ``qubit-virtual qubit'' decomposition, we can infer the expression of the rank-two density matrix $\rho_{\mc{S}\mc{F}_{N-m}}$,
\begin{equation}
	\rho_{\mc{S}\mc{F}_{N-m}}=\begin{pmatrix}
		p && 0 && s^{N}\sqrt{pq} && s^{m}\sqrt{1-s^{2(N-m)}}\sqrt{pq} \\
		0 && 0 && 0 && 0\\
		s^{N}\sqrt{pq}&& 0 &&  qs^{2(N-m)} && qs^{N-m}\sqrt{1-s^{2(N-m)}}\\
		s^{m}\sqrt{1-s^{2(N-m)}}\sqrt{pq}&& 0 && qs^{N-m}\sqrt{1-s^{2(N-m)}} && q(1-s^{2(N-m)})
	\end{pmatrix}.
\end{equation}
The eigenvalues, $\mu_1$ and $\mu_2$, of the matrix $\rho_{\mc{S}\mc{F}_{N-m}}\tilde{\rho}_{\mc{S}\mc{F}_{N-m}}$, are
\begin{equation}
	\mu_1= pq\left(1-s^{m}\right)^{2}(1-s^{2(N-m)}), \ \ \mu_2= pq\left(1+s^{m}\right)^{2}(1-s^{2(N-m)}).
\end{equation}
From the above eigenvalues, and based on~\cref{con2,enof,kwm}, we get the general analytic expression for the Holevo bound,
\begin{equation}
	\rchi(\mc{S}:\check{\mc{F}}_m)= h\left(r_p\right)-h\left(t_{p,m}\right),
	\label{chi1}
\end{equation}
where
\begin{equation}
	r_p= \frac{1}{2}\left(1+\sqrt{1-4pq\left(1-s^{2N}\right)}\right) \ \ \text{and} \ \ t_{p,m}=\frac{1}{2}\left(1+\sqrt{1-4pq\left(s^{2m}-s^{2N}\right)}\right).
\end{equation}
From this expression, it is a simple exercise to show that for good decoherence (i.e. in the limit $s^{N-m} \ll  s^m$) we have the following simplification
\ba
\rchi(\mc{S}:\check \cF_m)= H_\cS +\frac 1 2 \log_{2}\left(pqs^{2m}\right) +\sqrt{1-4pqs^{2m}}\operatorname{Arctanh}_2\left(\sqrt{1-4pqs^{2m}}\right),
\ea
such that $H_{\mc{S}}=-p \log_2(p)-q \log_2(q)$ is the von Neumann entropy of the decohered state of the central qubit $\mc{S}$.

In the subsequent two sections, based on the above result, we will illustrate the following statements made in the manuscript. The first statement is that the redundancy evaluated through the quantum mutual information $\cI$ overestimates the redundancy from the viewpoint of observers eavesdropping on fragments $\cF$ of the environment via optimal measurements. The difference between the two quantities can be computed numerically. The second statement concerns the weak dependence of the re-scaled quantum mutual information and Holevo bounds on the state preparation of the central qubit $\cS$, i.e. on the probability of the pointer states ``$p$''.
 
\section{Redundancy}

\begin{figure}
	\includegraphics[width=0.58\textwidth]{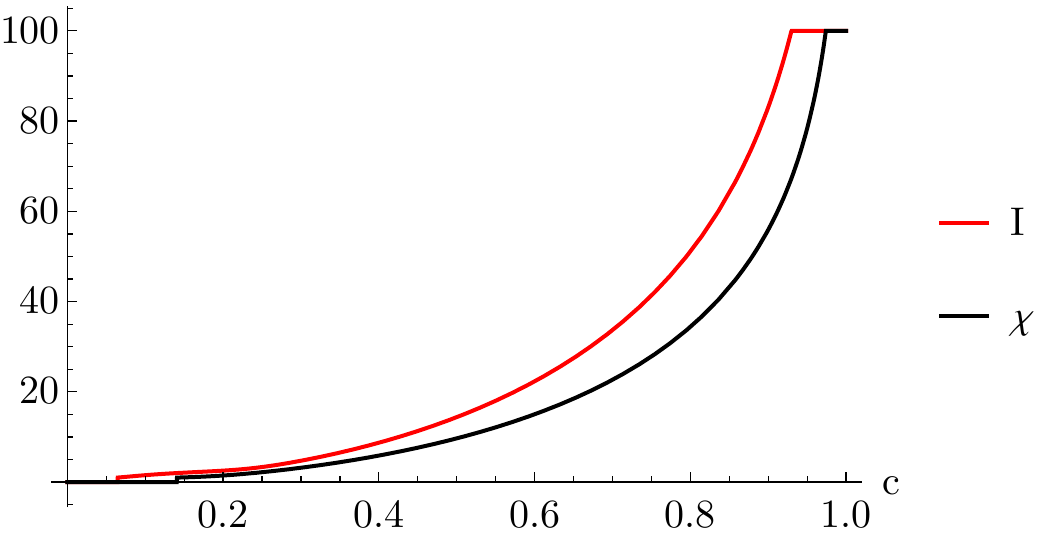}
	\caption{\label{Red}Plots of the redundancy $R_{0.1}$ computed through the mutual information $I$ (red curve) and the Holevo bound $\chi$ (black curve), as a function of $c$, for $p=1/2$ and $N=100$.}
\end{figure}

In~\cref{Red}, we illustrate the behavior of the redundancy $R_{\delta}=N/m_{\delta}$ (with information deficit $\delta=0.1$), evaluated using $I$ (red curve) and $\rchi$ (black curve), as a function of the parameter $c$. We observe that the redundancy is monotonically increasing as a function of $c$, such that the maximum redundancy is achieved when $c \rightarrow 1$. The latter is due to the fact that in the limit where $c \rightarrow 1$ the coupling between the central qubit and each qubit from the environment is now modeled by the {\tt c-not} gate, and the branching state is a GHZ state. We also observe that for almost all values of $c \in \ [0,1]$ the redundancy computed using the quantum mutual information is an overestimate to the redundancy observed by eavesdropping on fragments of the environment through optimal measurements.

\begin{figure}[h!]
	\centering
	\subfigure[]{
		\includegraphics[width=.485\textwidth]{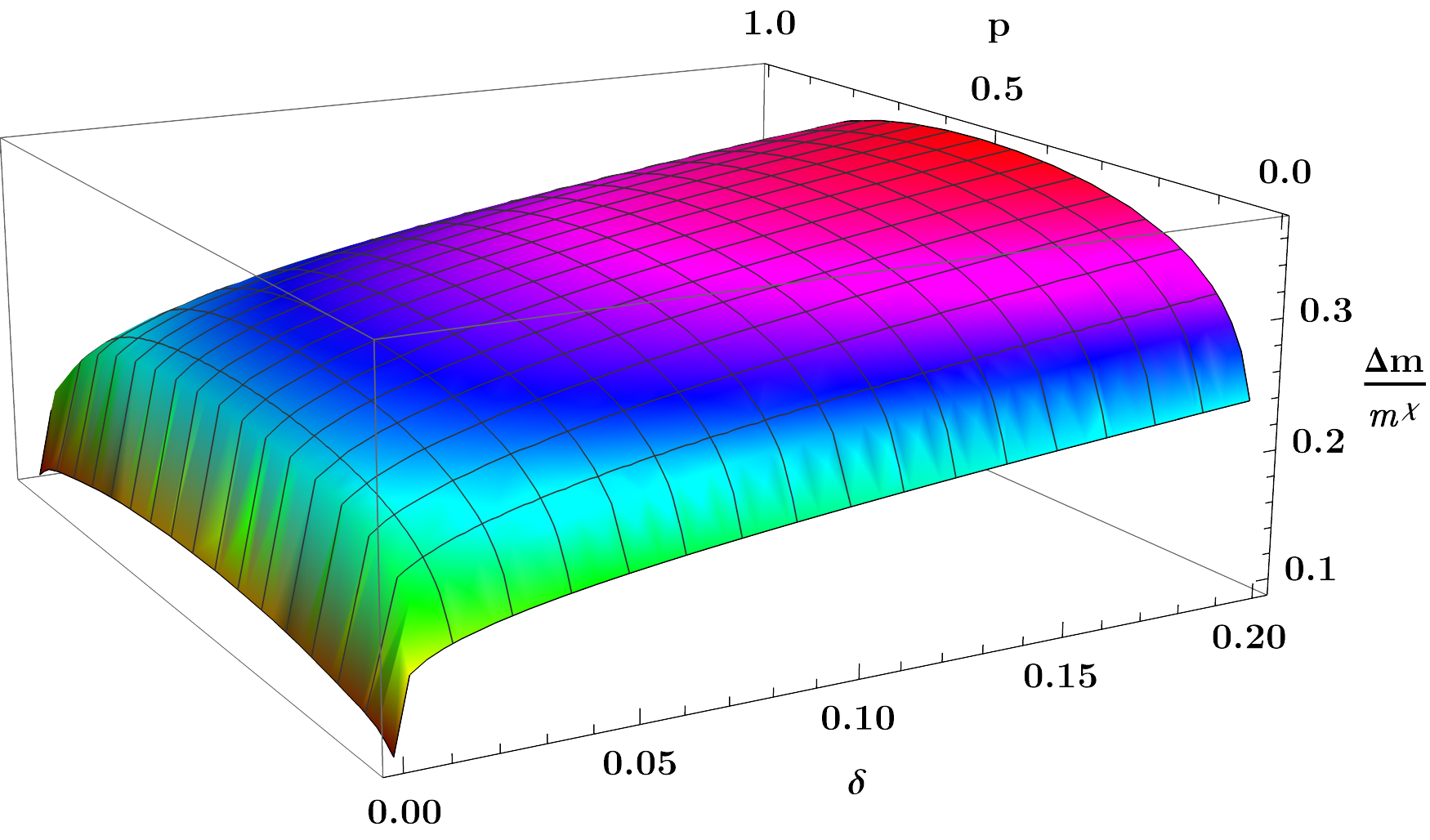}
	}
	\subfigure[]{
		\includegraphics[width=.485\textwidth]{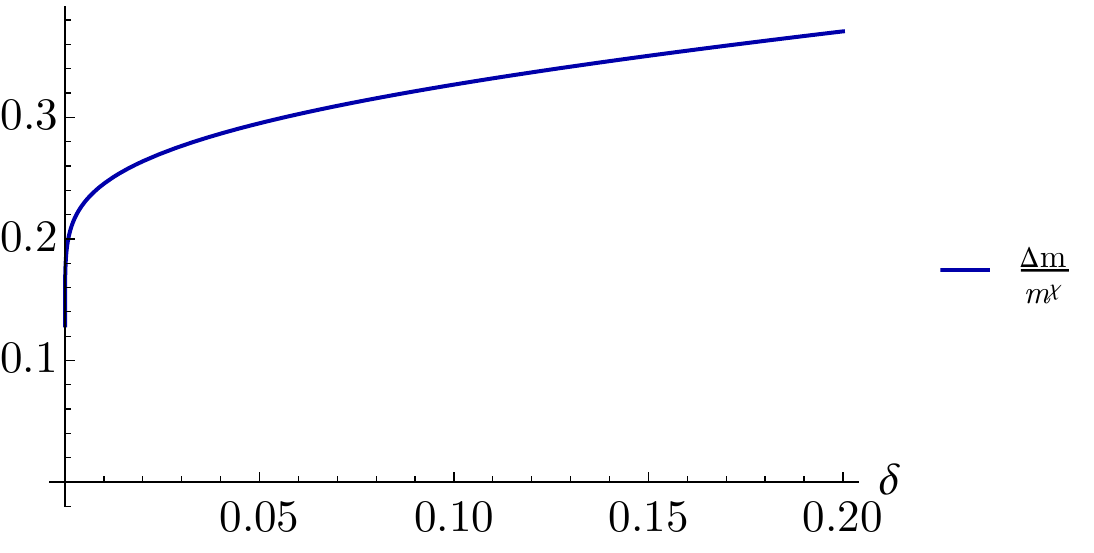}
	}
	\caption[]{\label{deltam2}Plots of $\Delta m /m^{\chi}$ as a function of $\delta \in [0,0.2]$ and $p$. In panel (a), we plot $\Delta m /m^{\chi}$ for all values of $p \in [0.01,0.99]$, and in panel (b) we focus on the case of $p=1/2$.}
\end{figure}

Using the analytic expressions of the mutual information and the Holevo bound, we can numerically determine the degree with which the redundancy obtained from the mutual information overestimates the actual redundancy. We can define $\Delta m= m^{\chi}-m^{I}$, such that $m^{\chi}$ ($m^{I}$) represents the minimum number of environmental qubits needed in order for the Holevo bound (mutual information) to reflect the missing information about the central qubit $\mc{S}$, up to an information deficit $\delta$, i.e., $(1-\delta)H_{\mc{S}}$. Therefore, the quantity $\Delta m/m^{\chi}$ reflects the difference between the redundancies obtained from the two different information theoretic measures ($I$ and $\chi$). In Fig.~\ref{deltam2}, we illustrate, in the limit of large $N$ ($N \rightarrow \infty$), $\Delta m/m^{\chi}$ as a function of $\delta$ and $p$. In panel (a), we observe that the maximum difference, for a given value of $\delta$, is attained for $p=1/2$. From panel (b), in the limit where $\delta \rightarrow 0$ we conclude that the redundancies differ by at most $\sim 13\%$, and for $\delta=0.2$ (similar to the case presented in the inset of Figure 1 of the manuscript~\cite{darwint1}) the difference is at most $\sim 37\%$.

\section{Breaking of universality}
\begin{figure}[h!]
	\centering
	\subfigure[]{
		\includegraphics[width=.6\textwidth]{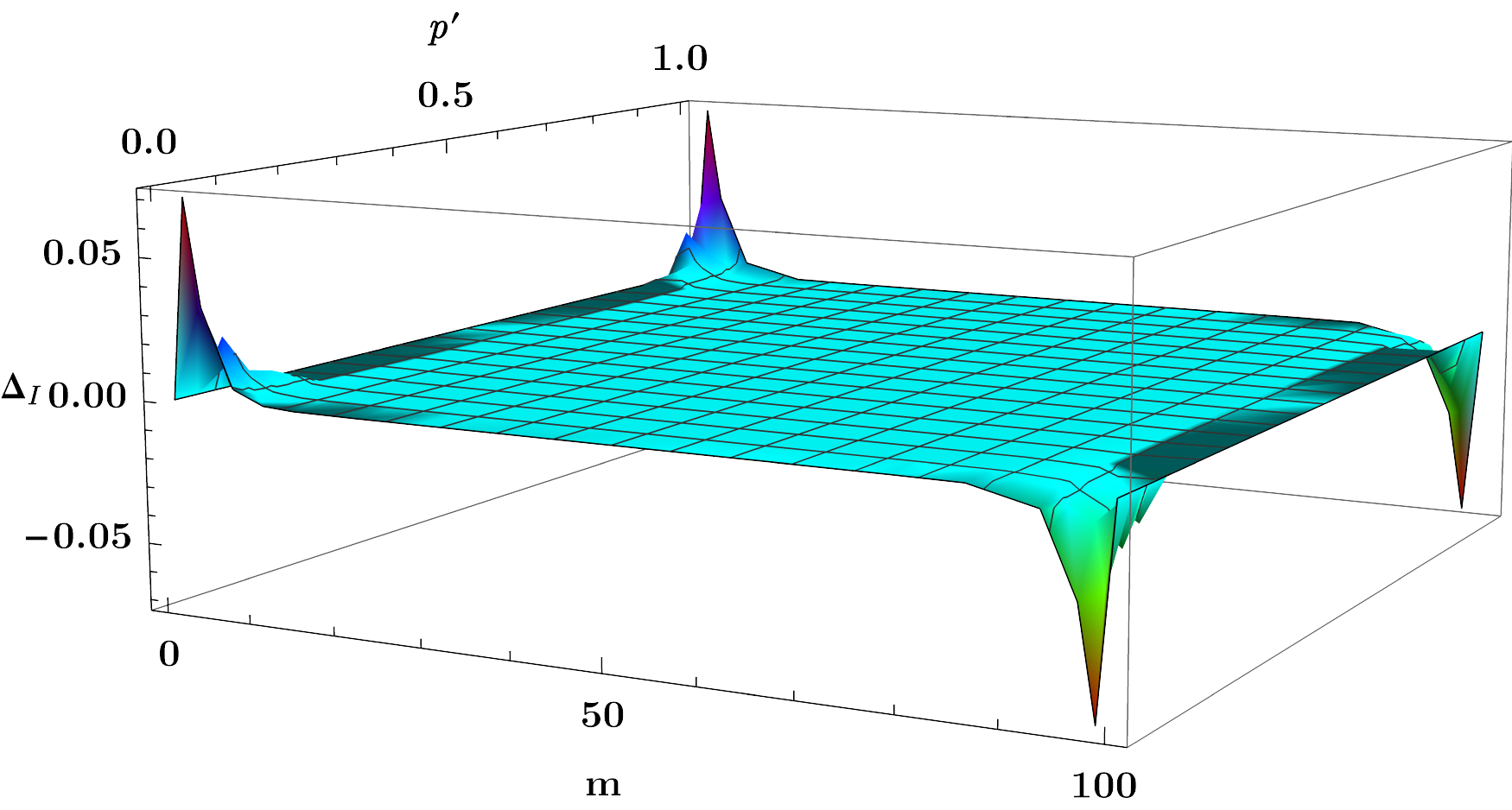}
	}
	\subfigure[]{
		\includegraphics[width=.6\textwidth]{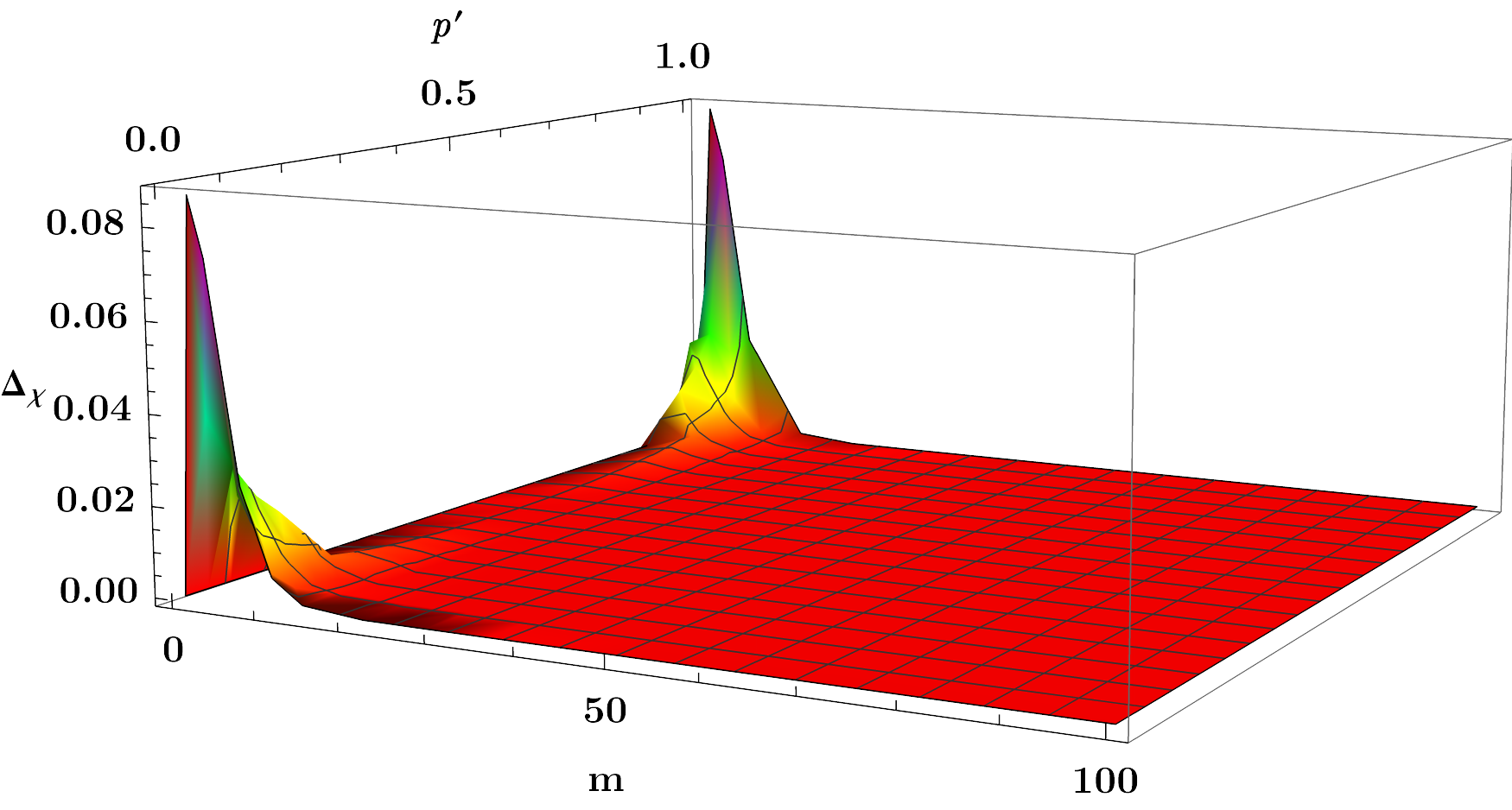}
	}
	\caption[]{\label{delta}Plots of $\Delta_{{I}}$ (panel (a)) and $\Delta_{\rchi}$ (panel (b)), as a function of $m$ and $p^{\prime}$, for $N=100$ and $c=\sqrt{0.4}$. Note that changing the value of the parameter $c$ for a fixed total number of qubits $N$ is equivalent to a straightforward re-scaling of the x-axis of our plots (displaying the number of qubits ``$m$'' in a typical environmental fragment), with $\log(s)$ as the re-scaling factor~\cite{darwint1}. It is also noteworthy that the initial rise of the mutual information or the Holevo bound is independent of the total number of qubits $N$ in our environment~\cite{darwint1}.}
\end{figure}

In the manuscript, we show that the quantum mutual information and the Holevo bound display a universal scaling behavior which is weakly dependent on the state preparation of the central qubit $\cS$. Namely, through appropriate re-scaling, the two information theoretic quantities exhibit a universal rising behavior insensitive to the value of the parameter $p$, up to some limitations. Here, our goal is to quantify and illustrate these limitations, which shows the breaking of the aforementioned universality when the parameter $p$ approaches a value of zero or one. To this end, we choose the case of the initial equal superposition ($p=1/2$) as a reference, and we compare its corresponding expressions of the mutual information and the Holevo bound with those for arbitrary $p^{\prime}$, which leads to defining $\Delta_{{I}}$ and $\Delta_{\rchi}$ as follows.
\begin{equation}
	\Delta_{{I}}={I}_{p=1/2}(\mc{S}:\mc{F}_m)-\frac{{I}_{p^{\prime}}(\mc{S}:\mc{F}_m)}{H_{p^{\prime}}(\mc{S})},
	\label{}
\end{equation}
such that ${I}_{p=1/2}(\mc{S}:\mc{F}_m)$ and ${I}_{p^{\prime}}(\mc{S}:\mc{F}_m)$ are the values of the mutual information for $p=1/2$ and arbitrary $p^{\prime}\in [0,1]$, respectively. The quantity $H_{p^{\prime}}(\mc{S})$ is the maximum von Neumann entropy of the system $\mc{S}$ corresponding to $p^{\prime}$ (i.e. the plateau of the curve of ${I}_{p^{\prime}}(\mc{S}:\mc{F}_m)$). Similarly, for the Holevo bound we define the quantity $\Delta_{\rchi}$ such that
\begin{equation}
	\Delta_{\rchi}=-\rchi_{p=1/2}(\mc{S}:\check{\mc{F}}_m)+\frac{\rchi_{p^{\prime}}(\mc{S}:\check{\mc{F}}_m)}{H_{p^{\prime}}(\mc{S})}.
	\label{}
\end{equation}
In~\cref{delta}, we plot $\Delta_{{I}}$ and $\Delta_{\rchi}$ as functions of both $m$ and $p^{\prime}$. For small $m$, the breaking of the universal rise of both the mutual information and the Holevo bound is observed when $p^{\prime}$ is either close to zero or one, otherwise $\Delta_{{I}} \approx 0$ and $\Delta_{\rchi} \approx 0$. For completeness, we also plot our quantities for large values of $m$ (i.e. when the environment fraction approaches one). In this limit, $\Delta_{\rchi} = 0$ for all $p^{\prime}$ while $\Delta_{{I}} \neq 0$ for $p^{\prime}$ close to zero or one. This is due to the fact that the quantum mutual information rises beyond the plateau when we capture almost all of the environment, which is a direct consequence of the purity of the state of the universe $\mc{S}\mc{E}$.

\section{Connection to the photon model}
As mentioned towards the end of the manuscript~\cite{darwint1}, our results naturally extend to the realistic model of a photon environment, as studied in Refs.~\cite{Riedel2010PRL,Riedel2011NJP}. In fact, the quantum mutual information expression is the same as the one derived in the photon scattering model. More specifically, if we consider a dielectric sphere, as our system of interest $\mc{S}$, initially in a spatial superposition such that $| \psi_{\mc{S}} \rangle =\sqrt{p} \delta\left(\vec{x}-\vec{x}_{1}\right)+\sqrt{q} \delta\left(\vec{x}-\vec{x}_{2}\right)$. The environment is composed of $N$ photons, originally emitted from a point source (for simplicity we assume point source illumination). In this case, it was shown~\cite{Riedel2010PRL,Riedel2011NJP} that the quantum mutual information has the following form,
\begin{equation}
	I (\mathcal{S}: \mathcal{F}_m)=1+\frac{1}{\log(2)}\sum_{i=1}^{\infty} \frac{\Gamma^{(1-f) i}-\Gamma^{f i}-\Gamma^{i}}{2 i(2 i-1)},
	\label{phot}
\end{equation}
where $\Gamma^{x}= (p-q)^{2}+4pq \exp(-tx/\tau_D)$ such that $x \in \{1,f,1-f\}$, $\tau_D$ is the decoherence time, and $f=m/N$ is the fraction of photons we access. The above expression is in direct agreement with our the results~\cite{darwint1}. In particular, from the specific form of equations (5-7) of the manuscript we can compute the analytic expression of the von Neumann entropies of the states $\rho_{\mc{S}}$, $\rho_{\mc{F}_m}$, and $\rho_{\mc{S}\mc{F}_m}$,
\begin{equation}
	H_{\mc{S}}=-\lambda_{-}\left(N, p\right) \log_{2} \left(\lambda_{-}(N,p)\right)-\lambda_{+}(N,p) \log_{2} \left(\lambda_{+}(N,p)\right),
\end{equation}
\begin{equation}
	H_{\mc{F}_m}=-\lambda_{-}(m,p) \log_{2} \left(\lambda_{-}(m,p)\right)-\lambda_{+}(m,p) \log_{2} \left(\lambda_{+}(m,p)\right),
\end{equation}
\begin{equation}
	H_{\mc{S}\mc{F}_m}=-\lambda_{-}(N-m,p) \log_{2} \left(\lambda_{-}(N-m,p)\right)-\lambda_{+}(N-m,p) \log_{2} \left(\lambda_{+}(N-m,p)\right),
\end{equation}
with $\lambda_{\pm}(k,p)=\frac{1}{2}\left(1 \pm \sqrt{\left(q-p\right)^{2}+4s^{2k}pq}\right)$. Generally, we have, for $|x|<1$,
\begin{equation}
	\log_{2}(1+x)=\frac{1}{\log(2)}\sum_{i=1}^{\infty}(-1)^{i+1} \frac{x^{i}}{i}, \ \ \text{and} \ \ \log_{2}(1-x)=-\frac{1}{\log(2)}\sum_{i=1}^{\infty} \frac{x^{i}}{i}.
\end{equation}
Therefore,
\begin{equation}
	-(1-x)\log_{2}(1-x)-(1+x)\log_{2}(1+x)=-\frac{2}{\log(2)}\sum_{i=1}^{\infty} \frac{x^{2i}}{2i(2i-1)}.
\end{equation}
It is, then, a straightforward exercise to get
\begin{equation}
	I (\mathcal{S}: \mathcal{F}_m)=1+\frac{1}{\log(2)}\left(\sum_{i=1}^{\infty} \frac{X_{N-m}^{2i}-X_{m}^{2i}-X_{N}^{2i}}{2i(2i-1)}\right),
	\label{series}
\end{equation}
such that
\begin{equation}
	X_i=\sqrt{\left(p-q\right)^2+4s^{2i}pq} \leq 1.
\end{equation}
Comparing this expression of the quantum mutual information with that of the photon model (cf.~\myeqref{phot}) we recognize the exact similarity such that $s^{N} \equiv \exp(-t/2\tau_D)$. Therefore, the universal rise (one of our main results) of the quantum mutual information also applies to the realistic photon scattering model.

\bibliography{opm2}